\documentclass[12pt]{iopart}

\usepackage{graphicx}

\newcommand{\bp}{\mathbf{p}}

\newcommand{\bv}{\mathbf{v}}

\newcommand{\bR}{\mathbf{R}}
\newcommand{\bbr}{\mathbf{r}}

\newcommand{\sep}{ \ \ \ , \ \ \ }

\newcommand{\beq}{\begin{equation}}
\newcommand{\eeq}{\end{equation}}
\newcommand{\beqn}{\begin{eqnarray}}
\newcommand{\eeqn}{\end{eqnarray}}
\newcommand{\pp}{\partial}
\newcommand{\dd}{{\rm d}}
\newcommand{\ee}{{\rm e}}

\newcommand{\eq}{Eq.\ }
\newcommand{\eqs}{Eqs\ }
\newcommand{\fig}{Fig.\ }

\newcommand{\la}{\langle}
\newcommand{\ra}{\rangle}

\newcommand{\ce}{{\rm ce}}

\begin{document}

\title[Active particles under confinement]{Active particles under confinement: Aggregation at the wall and gradient formation inside a channel}

\author{Chiu Fan Lee}

\address{Department of Bioengineering, Imperial College London
\\
South Kensington Campus, London SW7 2AZ, U.K.}
\ead{c.lee@imperial.ac.uk}
\begin{abstract}
I study the confinement-induced aggregation phenomenon in a minimal model of 
 self-propelled particles inside a channel.  Starting from first principles, I  derive a set of equations that govern the density profile of such a system at the steady-state, and calculate analytically how the aggregation at the walls varies with the physical parameters of the system. I also investigate how the gradient of the particle density varies  if the inside of the channel is partitioned into two regions within which the active particles exhibit distinct levels of fluctuations in their directions of travel. 
\end{abstract}
\submitto{\NJP}
\maketitle

\section{Introduction}

Aggregation of motile organisms, or self-propelled particles in general, is a familiar phenomenon, which is currently driving an intense interest in many scientific disciplines that include physics, biology and sociology \cite{here}. Many different physical mechanisms can lead to the aggregation of active systems. For example, these mechanisms can be directional alignment interactions \cite{peruani_pre06,
Ginelli_prl10,chate_prl06,narayan_science07,gregoire_prl04}, density-dependent motility \cite{peruani_prl11,farrell_prl12}, and aggregation at the wall 
when the system is under confinement \cite{kudrolli_prl08,wensink_pre08,kaiser_prl12,costanzo_jpcm12}. In this theoretical paper, I will  focus on the latter mechanism: confinement-induced aggregation of active particles. Experimentally, this mechanism is of particular interest given its inevitability due to the finite size of an experimental apparatus. The advent of microfluidic devices as an experimental tool 
to study active systems further motivates the current study since the channel walls constitute an integral part of the system \cite{ebbens_softmatter10}.

 \begin{figure}[b]
 \label{pic}
 \begin{center}
 \includegraphics[scale=.5]{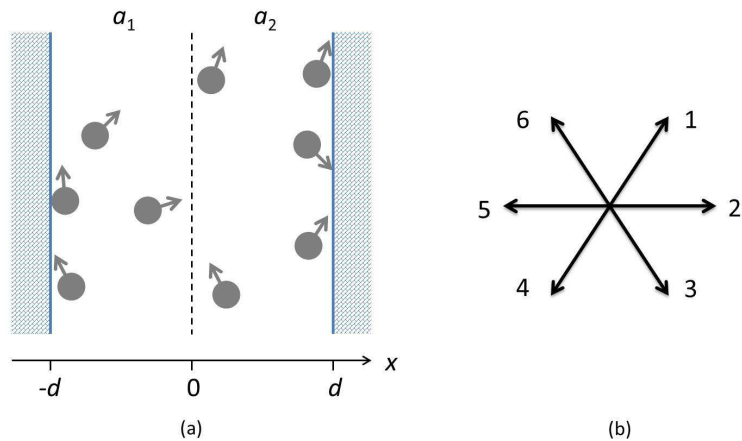}
 \caption{(a) A schematic depiction of the model system considered in this paper. Active particles are confined in a channel that is  infinite  in the $y$-direction but bounded in the $x$-direction. The walls are assumed to be perfectly rigid in that the particles cannot penetrate them but the particles are free to rotate at and slide along the walls. The channel inside is allowed to have two distinct rotational diffusion constants: $a_1$ for $x\geq 0$ and $a_2$ for $x<0$. (b) The simplified model also considered here  restrict the active particles to have only six distinct directions of travel. 
 }
 \end{center}
 \end{figure}

In the classical theory of liquids, aggregation of particles close to the wall is 
well understood and results from  the pairwise interactions, such as volume exclusion, of the particles in the system (e.g., see \cite{hansen_b} for a review). Density functional theory is an effective way of investigating such a phenomenon, and  
 a parallel effort has been pursued recently for a system of self-propelled rods  \cite{wensink_pre08}. On the other hand, if pairwise interactions between the particles become negligible, such a confinement-induced aggregation phenomenon  will cease to exist in an equilibrium system, while it will continue to persist in an active system.
 This is the effect I will focus on here. Specifically, I will 
 investigate a system of active particles  that  travel at a fixed speed in a channel, undergo rotational diffusion, and react only with the walls.
 I will derive from first principles a set of equations that govern the density profile of this system, and show that the density function can be written as a series expansion of the Mathieu functions  \cite{mclachlan_b51}. To make further analytical process, I will introduce a simplified model in which the particles' directions of travel are discretised.

 Besides determining quantitatively how confinement induces aggregation, I will also discuss the gradient formation of particle density in a channel that is partitioned into two regions within which the active particles exhibit distinct rotational fluctuations (\fig \ref{pic}(a)).
  Such a scenario can for instance be engineered by having two background mediums coexisting in a microfluidic channel, and has been recently investigated in a combined experimental and numerical work \cite{ebbens_softmatter12}.

The structure of this paper is as follows: I will first specify the model under study in Section \ref{model}, and  present three ``predictions'' based on intuitive arguments in Section \ref{scaling}. The mathematical formulation of the system will be discussed  in Section \ref{sec:math}. A simplified model and its full analysis will be presented in Section \ref{sec:simple}. The paper ends with a discussion of the findings.

\section{Model}
\label{model}
I consider a minimal model of active particles in two dimensions, where every particle is assumed to have constant 
speed $u$. 
Noise, of strength $a$, is incorporated in the direction of travel and the particles interact only with the boundary of the system. 
Mathematically,  the equations of motion for such a $N$-particle system are
\beqn
\label{eq:rr}
\frac{\dd  \bbr_i}{\dd t} &=& u \bv(\theta_i)
\\
\label{eq:theta}
\frac{\dd  \theta_i}{\dd t} &=& \sqrt{2a}\eta_i(t)
\eeqn
where $1 \leq i \leq N$, $\bR \equiv (\bbr_1, \ldots, \bbr_N)$, $\Theta \equiv (\theta_1, \ldots, \theta_N)$, $\bv(\theta)\equiv (\cos \theta, \sin \theta)$, and (\eq \ref{eq:theta}) is an It\^{o} stochastic differential equation \cite{gardiner_b} such that
\beq
\la \eta_i(t) \ra =0 \sep \la \eta_i(t)\eta_j(t') \ra = \delta_{ij} \delta(t-t') \ .
\eeq
The specific geometry considered is depicted in \fig \ref{pic}(a) where the walls are assumed to be perfectly rigid and smooth such that the particles are free to rotate at and slide along  the walls.
Note that in this model, I ignored the thermal translational diffusion $D$ of the particles \cite{peruani_prl07, romanczuk_prl11}. As explained in \cite{pototsky_epl12}, this simplification is valid if $D \ll u^2 a^{-1}$, which applies, e.g., for the bacterium {\it Escherichia coli}.

Without loss of generality, I will from now on set the length scale and temporal scale so that the speed $u$ is one and the width of the channel $2d$ is 2, although I will continue to keep the symbols in the equations so that the dimensions of the terms in the equations are always clear.

\section{Intuitive arguments}
\label{scaling}
I will first analyse the system based on physical arguments in this section. Since there are no interactions between particles by assumption, I can focus on the average behaviour of a single particle. Specifically, denoting the average time the average particle spent at the wall by $\tau_w$, and the average time spend in the channel by $\tau_c$, the boundary condition suggests that if  a particle is stuck at a wall, i.e., having its direction of travel pointing towards the wall, then it will only leave the wall if it manages to rotate its direction to point away from it, so the average time spent at the wall can be estimated as  the time it takes to rotate, say, by one radian. As a result,
\beq
\label{eq:tauw}
\tau_w \sim 1/a \ .
\eeq
To estimate $\tau_c$, one has to worry about whether the particles in the channel is dominantly ballistic or diffusive. These two behaviours are dictated by the 
rotational diffusion constant $a$. In other words, if $a$ is large, then the particles will have rotated around many times over before reaching the walls
and so the motion of the particle is diffusive with an effective diffusion constant $D_{eff} = u^2/a$ (see, e.g., Ch.~6 in \cite{berg_b}).
On the other hand, if $a$ is small, the motion of the particle inside the channel will be predominantly ballistic. Therefore, 
\beq
\tau_c \sim
\left\{ \begin{array}{ll}
d^2/D_{eff} \sep & {\rm diffusive}
\\
d/u \sep & {\rm ballistic}
\end{array}
\right.
\eeq
Remembering that the units are set such that $d=u=1$,  $\tau_c$ becomes
\beq
\tau_c \sim
\left\{ \begin{array}{ll}
a \sep & {a \gg 1}
\\
1 \sep & {a \ll 1}
\end{array}
\right.
\eeq
Together with (\eq \ref{eq:tauw}), one may estimate the ratio of the number of particles inside the channel $n_c$ over the number of particles at the walls $n_w$ to be
\beq
\frac{n_c}{n_w} \sim \frac{\tau_c}{\tau_w}
\sim
\left\{ \begin{array}{ll}
a^2 \sep & {a \gg 1}
\\
a \sep & {a \ll 1} \ .
\end{array}
\right.
\eeq
\vspace{.15in}

\noindent
If the channel is now partitioned into two regions  system with distinct rotational diffusion constants $a_1$ and $a_2$ (\fig \ref{pic}(a)), an interesting question that is of experimental interests  \cite{ebbens_softmatter12} is which side of the region will be of higher density. Focusing on the $a_1,a_2 \gg 1$ regime, one may expect that the particles would behave diffusively inside the channel, with effective diffusion constant $D_1\sim 1/a_1$ and $D_2\sim 1/a_2$ in the two distinct regions. Intuitively, if there is a gradient in the system, then the region with the lower diffusion constant would have a higher number of particles since the slow mobility would serve to localise particles  \cite{peruani_prl11,farrell_prl12}.  
Interestingly, such an effect has also been observed in the embryo of the round worm {\it Caenorhabditis elegans}, where regions with lower diffusion constants serve to localise certain diffusing proteins 
 \cite{daniels_dev10,griffin_cell11}.  If this expectation extends to our active system, then one prediction would be that for $a_1,a_2 \gg 1$, $a_1 < a_2$ implies $n_c^{(1)}<n_c^{(2)}$ where $n_c^{(i)}$ denotes the number of particles in region $i$.

Summarising this section, the following three predictions seem to follow from the intuitive arguments presented above:
\begin{enumerate}
\item
In the symmetric case ($a_1=a_2=a$), if $a\ll 1$, then $n_c/n_w \sim a$ (?)
\item
In the symmetric case, if $a\gg 1$, then $n_c/n_w \sim a^2$ (?)
\item
In the asymmetric case, if $a_1<a_2$ and $a_1,a_2 \gg 1$, then $n_c^{(1)} < n_c^{(2)}$ (?)
\end{enumerate}
With the mathematical tools developed in the next section, I will show that in fact only the first ``prediction'' is true.

\section{Mathematical formulation}
\label{sec:math}
I will now provide a mathematically rigorous formulation of the problem. 
If we denote the probability distribution of the density of particles in the state $(\bR, \Theta)$ at time $t$ by $f(t, \bR, \Theta)$, then the Fokker-Planck equation 
corresponding  to the system is \cite{Zwanzig_b}:
\beq
\label{feq}
\frac{\pp f}{\pp t} =  \sum_i \bigg\{a \frac{\pp^2 }{\pp \theta_i^2}f  -u\nabla_{\bbr_i} \cdot [\bv(\theta_i) f] \bigg\}
 \ .
\eeq
Since by assumption, the particles do not interact with each other, the dynamics of the system is fully captured by the corresponding single-particle density function $p(\bbr_1, \theta_2) \equiv N \int \dd \bbr_2 \cdots \bbr_N \dd \theta_2 \cdots \theta_N f(\bR, \Theta)$, which is governed by the following dynamical equation  \cite{huang_b,lee_pre10}:
\beq
\label{eq:main}
\frac{\pp p}{\pp t} = a \frac{\pp^2 p}{\pp \theta^2}  -u\nabla_{\bbr_i} \cdot [\bv(\theta)p] 
 \ .
\eeq
Focusing on the steady-state solution and assuming the channel is infinite along the $y$-direction, the equation simplifies to
\beq
\label{eq:ma}
 a\frac{\pp^2 p}{\pp \theta^2}  = u\cos \theta \frac{\pp p}{\pp x} 
 \ .
\eeq
Employing the separation of variables method by assuming that  $\rho(x,\theta)=A(\theta)B(x)$, I arrive at the two equations
\beqn
\label{eq:a}
\pp_{\theta \theta}A &=& q \cos \theta A \ ,
\\
\label{eq:b}
\pp_x B &=& qa B
\ ,
\eeqn
where $q$ corresponds to a set of constants to be determined. \eq (\ref{eq:b}) can be easily solved:
\beq
B= K\ee^{Dq x}
\ ,
\eeq
while \eq (\ref{eq:a}) indicates that the function $A$ is a Mathieu function, usually denoted as $\ce_{2n}(\theta,q_{2n})$
where $n\in {\bf N}$, such that the corresponding characteristic number is zero  \cite{mclachlan_b51}. Since the characteristic numbers for these types of Mathieu functions are identical for $\pm q$, the expansion for $p$ is
\beq
\label{eq:mathieu}
p(x,\theta)= \sum_{n\geq 0} \left[a_n\ee^{q_{2n}ax} \ce_{2n} (\theta/2, 2q_{2n})
+b_n\ee^{-q_{2n}ax} \ce_{2n} (\theta/2, -2q_{2n})\right]
\ .
\eeq
As usual, the expansion parameters are obtained by the boundary conditions, which I will determine now.

\begin{figure}[t]
\begin{center}
\includegraphics[scale=.63]{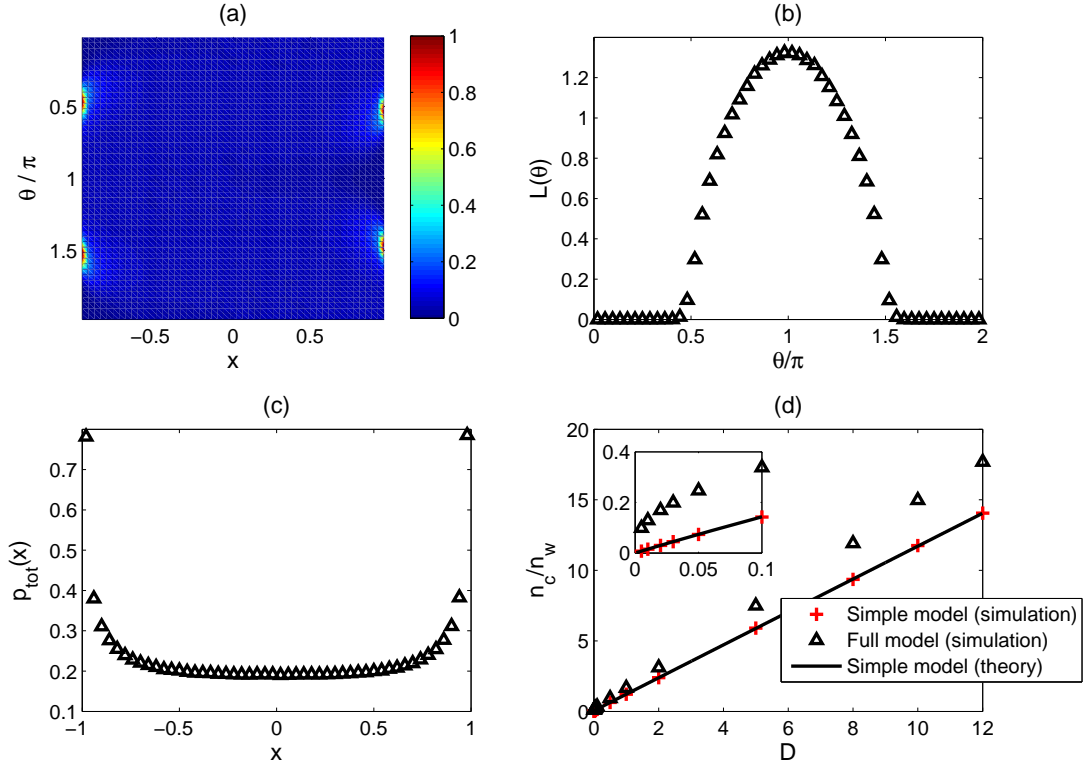}
\end{center}
\caption{Steady-state distribution of active particles in the  symmetric case ($a_1=a_2$).
(a) The density map of the active particles in the system at the steady state as a function of position and direction based on extensive Brownian dynamics simulation \cite{BDS}. The highest density is arbitrarily set to one (see colorbar). (b) The normalised distribution of the particle's direction at the left wall $L(\theta)$. (c) The normalised density of particles within the channel $p_{tot}(x) =\int_0^{2\pi} \dd \theta p(x,\theta)$. (d) The ratio of the number of particles in the channel {\it vs}.\ that at the wall ($n_c/n_w$) as a function of $a$, for both the full and simplified 6-direction models, the curve depicts the analytical expression in \eq (\ref{anal}).
}
\label{fig2}
\end{figure}

Denoting the number of particles at the wall on the left ($x=-d$) travelling in the direction $\theta$ by $L(\theta)$,   $L(\theta)$ must be zero for $-\pi/2 < \theta <\pi/2$ as particles travelling in those directions would not stay at the wall (\fig \ref{fig2}(b)). For $ \pi/2<\theta <3\pi/2$, the equation for $L(\theta)$ can be derived by the conservation of particles as follows: The flux of particles that arrive at the wall is $-u \cos \theta p^L(\theta)$ where $p^L(\theta)$ denotes $p(x=-d, \theta)$. At the wall, the particles that are stuck (i.e., particles with direction $-\pi/2<\theta <\pi/2$) undergo rotational diffusion. Therefore, the density function $L(\theta)$ satisfies the following differential equation:
\beq
\label{eq:l}
a  \frac{\pp^2 L}{\pp \theta^2} =u\cos \theta p^L(\theta) \sep \pi/2<\theta <3\pi/2
\ ,
\eeq
with the boundary conditions $L(\theta =\pm \pi/2)=0$. 
Finally, the flux of particles that leave the wall corresponds to the flux of particles that reach the angle $\theta = \pm \pi/2$. For instance, for the out-flux source at $\theta = \pi/2$, the magnitude of the out-flux is $|a \pp L(\theta=\pi/2) /\pp \theta |$. Together with the out-flux source at  $\theta =- \pi/2$, the overall out-flux at the left wall is
\beq
\label{eq:flux}
  a\left\{\delta(\theta-\pi/2) \left| \frac{\pp L(\theta= \pi/2)}{\pp \theta} \right|+\delta(\theta+\pi/2) \left| \frac{\pp L(\theta=-\pi/2)}{\pp \theta} \right|\right\}
  \ .
  \eeq

  Similarly, the equation governing the steady-state profile at the wall on the right $R(\theta)$ is
  \beq
  \label{eq:r}
  a  \frac{\pp^2 R}{\pp \theta^2} =-u\cos \theta p^R(\theta) \sep -\pi/2<\theta <\pi/2
  \ ,
  \eeq
  with the same boundary conditions: $R(\theta =\pm \pi/2)=0$
  
As aforementioned, at the steady-state, the particles that are accumulated at the walls eventually re-enter the channel at four delta sources (two at either wall) (see \fig \ref{fig2}(a)). To account for these influxes of particles from the walls,  \eq (\ref{eq:ma})  is modified as follows:
 \beq
 \label{eq:p}
  a\frac{\pp^2 p}{\pp \theta^2}  = u\cos \theta \frac{\pp p}{\pp x} +a\delta(x-d)\delta(\theta \pm \pi/2)  \left|   \frac{\pp R}{\pp \theta} \right| +a\delta(x+d)\delta(\theta \pm \pi/2)  \left|   \frac{\pp L}{\pp \theta} \right| 
  \ .
 \eeq
 In summary, the steady-state density profile of the active particles is obtained by solving the coupled differential equations \eqs (\ref{eq:l}), (\ref{eq:r}) and (\ref{eq:p}).

The standard way to proceed at this point is to substitute into the above coupled differential equations the expansion for $p$ in terms of the Mathieu functions (\eq (\ref{eq:mathieu})), and then try to ascertain the expansion coefficients accordingly. In practice, further analytical process is  difficult since manipulations of the Mathieu functions are  analytically intractable. Indeed, the numerical challenge to determine the characteristic numbers for Mathieu functions is comparable to performing Brownian dynamics simulation of the system \cite{mclachlan_b51,BDS}. 
Therefore, in order to  further reveal the underlying physics of this problem, I will now introduce and analyse a simplified version of the original model that admits full analytical solution.

\section{Simplified 6-direction model}
\label{sec:simple}
In this simplified model, the angular components are partitioned into six directions as shown in \fig \ref{pic}(b) and each particle will switch to a neighbouring direction of travel with rate $a$. I will denote the density function inside the channel travelling towards the $i$-th direction by $p_i$, and those at the wall on the right and on the left by $R_i$ and $L_i$ respectively.
The translational  symmetry along the $y$ direction indicates that $p_1=p_3$, $p_6=p_4$, and also that the functions $p$ only depend on $x$ but not $y$. As a result, 
the dynamical equations for the system are
\beqn
\pp_t p_1 &=& a(p_2+p_6-2p_1) -b\pp_x p_1
\\
\pp_t p_2 &=& 2a(p_1-p_2) -\pp_x p_2
\\
\pp_t p_5 &=& 2a(p_6-p_5) +\pp_x p_5
\\
\pp_t p_6 &=& a(p_1+p_5-2p_6) +b\pp_x p_6
\eeqn
where $b=\sqrt{3}/2$ since $b=\cos \pi/3$. In matrix form, 
$\dd \bp/\dd x = M \bp$
where $\bp = (p_1, p_2,p_5,p_6)^T$ and 
\beq
M=\left(
\begin{array}{cccc}
-2a/b & a/b & 0 & a/b
\\
2a & -2a & 0 &0
\\
0 & 0 & 2a & -2a
\\
-a/b & 0 & -a/b & 2a/b
\end{array}
\right)
\ .
\eeq
The four eigenvalues associated to the matrix $M$ are $\{0,0, -a\gamma/b, a\gamma/b\}$ where $\gamma = \sqrt{3+4b+4b^2}$, and the corresponding eigenvectors are the column vectors in the following matrix:
\beq
V=
\left(
\begin{array}{cccc}
0 & 1 & 2+\gamma-\frac{2b}{2b+\gamma} & 1
\\
0 & 1 & -4(b+b^2)-\frac{2b(2-\gamma)}{2b+\gamma} &\frac{2b}{2b+\gamma}
\\
0 & 1 & \frac{2b}{2b+\gamma} &-4(b+b^2)-\frac{2b(2-\gamma)}{2b+\gamma}
\\
0 & 1 & 1 & 2+\gamma-\frac{2b}{2b+\gamma}
\end{array}
\right)
\ .
\eeq
Furthermore, the steady-state solution is of the form:
\beq
\label{eq:exp}
p_h(x) = \sum_k c_k \ee^{\sigma_k x} V_{hk}
\ ,
\eeq
where $\sigma_k\in \{0,0, -a\gamma/b, a\gamma/b\}$  are the corresponding eigenvalues of $M$. The above expression is to be compared to Eq.~(\ref{eq:mathieu}) in the continuum model.

To determine the coefficients $c_k$ in Eq.~(\ref{eq:exp}), I employ the discrete version of the boundary conditions as discussed in the previous section (\eq (\ref{eq:l})), which for the wall on the left gives,
\beqn
L_1 &=&L_2 =0
\\
L_5 &=& \frac{u}{a} \left( p_5^L+bp_6^L\right)
\\
\label{eq:l6}
L_6 &=& \frac{u}{a} \left( \frac{p_5^L}{2}+bp_6^L\right)
\ ,
\eeqn
and the flux of particles out of the wall on the left is (see \eq (\ref{eq:flux}))
\beqn
p^L_1 &=&  \frac{aL_6}{bu} 
\\
\label{eq:cond}
& =&  \frac{p_5^L}{2b}+p_6^L \ ,
\eeqn 
where the second equality follows form \eq (\ref{eq:l6}).
Similar expressions can be readily found for the $R_i$.
In particular, the condition in \eq (\ref{eq:cond}) and the corresponding one for $p^R_i$ enable me to determine the coefficients in the expansion in \eq (\ref{eq:exp}) completely.

\begin{figure}[t]
\begin{center}
\includegraphics[scale=.63]{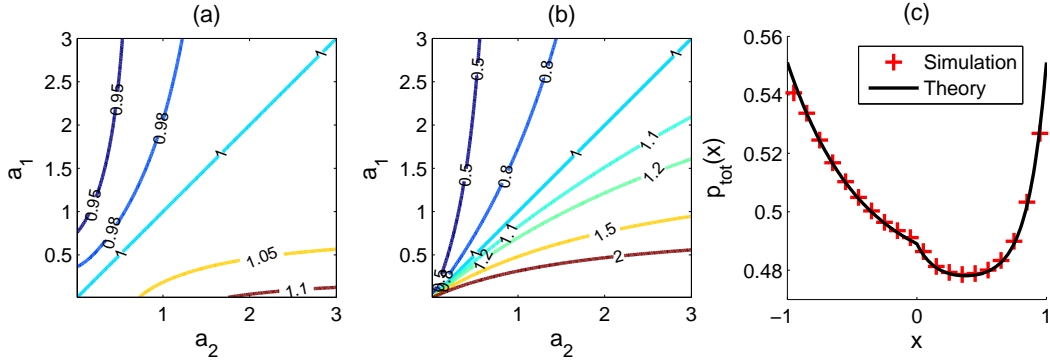}
\end{center}
\caption{Steady-state properties of the simplified system in the asymmetric case where two regions with distinct rotational diffusion constants co-exist. The results are based on the analytical theory presented in the main text. (a) The contour plot of the ratio of number of particles inside the left-half of the channel {\it vs}.\ the right-half of the channel excluding the particles at both walls $\left(n_c^{(1)}/n^{(2)}_c \right)$ as a function of $a_1$ and $a_2$. (b) The contour plot of the ratio of number of particles inside the left-half of the channel {\it vs}.\ the right-half of the channel including the particles at both walls  $\left((L_{tot}+n_c^{(1)})/(R_{tot}+n_c^{(2)})\right)$ as a function of $a_1$ and $a_2$. (c) The gradient of particle density ($p_{tot}(x)= \sum_kp_k(x)$) formed inside the channel for the case $a_1=0.2$ and $a_2=5$.
}
\label{fig3}
\end{figure}

\subsection{Symmetric case ($a_1=a_2=a$)}
In the symmetric case, the interesting quantity to investigate is the ratio of the number of particles inside the channel $n_c$ {\it vs}.\ the number of particles at the wall $n_w$.  The analytical expression for $n_c/n_w$ can be obtained straight-forwardly with the help of Mathematica \cite{mathematica}. Although the analytical expression is complicated and difficult to decipher, it does provide the following asymptotic expressions with respect to $a$:
\beq
\label{anal}
\frac{n_c}{n_w} =
\left\{
\begin{array}{ll}
\frac{4}{3 b} \ a \simeq 1.54a \sep & a \ll 1
\\
\frac{12 a \left(-1+\gamma+2 b \gamma+\gamma^2\right)}{\gamma \left(7+12 b^2+4 \gamma+2 b (7+3 \gamma)\right)}\ a \simeq 1.17a \sep & a \gg 1
\end{array}
\right.
\ .
\eeq
This result indicates that the ratio increases linearly in both asymptotic regimes, albeit with two different slopes. These analytical results are in perfect agreement with results from Brownian dynamics simulation (\fig \ref{fig2}(d)). Referring to Section \ref{scaling}, expectation (i)  is thus verified while expectation (ii) is invalidated. I will come back to discuss why this is the case in the Discussion Section.

\subsection{Asymmetric case ($a_1 \neq a_2$)}

In the two-region case as depicted in \fig \ref{pic}(a), the expansion coefficients in \eq (\ref{eq:exp}) are obtained by employing the boundary conditions at both walls (\eq (\ref{eq:cond})), and by matching the densities $p_i$ at $x=0$. Note that the matching condition at $x=0$ ensures the continuity of the density functions but not their slopes. This is fully supported by  simulation result (\fig \ref{fig3}(c)). 

 \fig \ref{fig3}(c) also shows that there is a gradient of particle density  inside the channel, one natural question is  which region possesses more particles. The answer can again be obtained from the analytical solution for $p_i$, $R_i$ and $L_i$, and the result is that the region with a larger rotational diffusion constant always has a smaller number of particles, whether the comparison concerns the particles inside the channel only (\fig \ref{fig3}(a)) or includes the particles at the walls (\fig \ref{fig3}(b)). In other words, if $a_1 <a_2$, then  $n_c^{(1)} > n_c^{(2)}$. Hence, the expectation (iii) discussed in Sect.~\ref{scaling} is not valid. Note that this surprising phenomenon has also been observed previously in a simulation study of a similar type of active system \cite{ebbens_softmatter12}.

\section{Discussion}
As we have seen, among the three expectations arisen from intuitive arguments in Sect.~\ref{scaling}, only the first expectation is verified. The reason that expectation (ii) is invalid is due to the overall conservation in the number of particles in the system. In other words, although it remains true that as $a\rightarrow \infty$, $n_w \sim 1/a$, $n_c$ cannot scale with $a$ as this would lead to a divergence in the total number of particles. So as $a$ increase, $n_c$ also increases but eventually saturates to a number bounded below by  the total number of particles in the system. This is why $n_c/n_w$ remains linear in $a$ even in the large $a$ regime. 

In the asymmetric case, in contradiction to the expectation (iii) from Sect.~\ref{scaling}, a higher $a_2$ in relation to $a_1$ always leads to a higher density of particles in region 1, irrespective of whether $a_1,a_2 \gg 1$ or not (see Fig.~\ref{fig3}(a)). A physical explanation may be that first, a higher $a$ leads to a lower accumulation of particles at the wall (Eq.~\ref{eq:tauw}); and second,
the exponent (i.e., the eigenvalues) scales with $a$ (\eq (\ref{eq:exp})  and so a higher $a$ will lead to more rapid decay of density from the wall (\fig \ref{fig3}(c)). These two complimentary effects turn out to be dominant in determining the gradient of the particle density, and thus lead to the observation that a higher $a$ always leads to a lower density of particle in that region. 
Of course, the validity of this argument is only confirmed by previous analytical investigation. 

In summary, I have studied the confinement-induced aggregation phenomenon in a  system of 
 self-propelled particles inside a channel. The model system under consideration is drastically simplified in order to focus on the effects of confinement on aggregation. In particular, the particles are assumed to be non-interacting and the walls are assumed to be perfectly rigid and smooth. Even in this simplified scenario, solving the rotationally-continuous  model is mathematically difficult and thus forbids an analytical understanding of the system. Therefore, I introduced a further-reduced model where the particles can only take six different directions of travel. This enabled me to solve the model analytically and to prove that when the rotational fluctuation $a$ is uniform in the system, the ratio of the number of particles inside the channel {\it vs}.\ that accumulated at the walls increases linearly with $a$ in both the $a\ll 1$ and $a \gg 1$ regimes. I also demonstrated that if one side of the channel has a larger rotational fluctuation, this inevitably leads a lower density of particles on that side. Given the fundamental nature of the model and the general nature of the observations, the effects of confinement on active systems discussed here should continue to be present in more elaborate models and in real experimental systems. 
 In particular, it would be interesting to study to what extend confinement-induced gradient formation in the bulk of an active system can lead to the emergence of pattern formation in the whole system.

\section*{References}

\begin{thebibliography}{10}


\bibitem{here}
See, e.g., this volume. 

\bibitem{peruani_pre06}
Fernando Peruani, Andreas Deutsch, and Markus B\"{a}r.
\newblock Nonequilibrium clustering of self-propelled rods.
\newblock {\em Physical Review E}, 74:030904, 2006.

\bibitem{Ginelli_prl10}
Francesco Ginelli, Fernando Peruani, Markus B\"{a}r and Hugues Chat\'{e}.
 \newblock Large-Scale Collective Properties of Self-Propelled Rods.
 \newblock {\em Physical Review Letters}, 104:184502, 2010.

\bibitem{chate_prl06}
Hugues Chat\'{e}, Francesco Ginelli, and Ra\'{u}l Montagne.
\newblock Simple model for active nematics: Quasi-long-range order and giant
  fluctuations.
\newblock {\em Physical Review Letters}, 96:180602, 2006.

\bibitem{narayan_science07}
Vijay Narayan, Sriram Ramaswamy, and Narayanan Menon.
\newblock Long-lived giant number fluctuations in a swarming granular nematic.
\newblock {\em Science}, 317:105--108, 2007.

\bibitem{gregoire_prl04}
Guillaume Gr\'{e}goire and Hugues Chat\'{e}.
\newblock Onset of collective and cohesive motion.
\newblock {\em Physical Review Letters}, 92:025702, 2004.

\bibitem{peruani_prl11}
Fernando Peruani, Tobias Klauss, Andreas Deutsch, and Anja Voss-Boehme.
\newblock Traffic jams, gliders, and bands in the quest for collective motion
  of self-propelled particles.
\newblock {\em Physical Review Letters}, 106:128101, 2011.

\bibitem{farrell_prl12}
F.~D.~C. Farrell, M.~C. Marchetti, D.~Marenduzzo, and J.~Tailleur.
\newblock Pattern formation in self-propelled particles with density-dependent
  motility.
\newblock {\em Physical Review Letters}, 108:248101, 2012.

\bibitem{kudrolli_prl08}
Arshad Kudrolli, Geoffroy Lumay, Dmitri Volfson, and Lev~S. Tsimring.
\newblock Swarming and swirling in self-propelled polar granular rods.
\newblock {\em Physical Review Letters}, 100:058001, 2008.
\newblock PRL.

\bibitem{wensink_pre08}
H.~H. Wensink and H.~L\"{o}wen.
\newblock Aggregation of self-propelled colloidal rods near confining walls.
\newblock {\em Physical Review E}, 78:031409, 2008.

\bibitem{kaiser_prl12}
A.~Kaiser, H.~H. Wensink, and H.~L\"{o}wen.
\newblock How to capture active particles.
\newblock {\em Physical Review Letters}, 108:268307, 2012.

\bibitem{costanzo_jpcm12}
A~Costanzo, R~Di Leonardo, G~Ruocco, and L~Angelani.
\newblock Transport of self-propelling bacteria in micro-channel flow.
\newblock {\em Journal of Physics: Condensed Matter}, 24:065101, 2012.

\bibitem{ebbens_softmatter10}
Stephen~J. Ebbens and Jonathan~R. Howse.
\newblock In pursuit of propulsion at the nanoscale.
\newblock {\em Soft Matter}, 6:726--738, 2010.

\bibitem{hansen_b}
Jean-Pierre Hansen and I.~R. McDonald.
\newblock {\em Theory of Simple Liquids}.
\newblock Academic Press, 2006.

\bibitem{mclachlan_b51}
N.W. McLachlan.
\newblock {\em Theory and application of Mathieu functions}.
\newblock Clarendon, 1951.

\bibitem{ebbens_softmatter12}
Stephen Ebbens, Gavin Buxton, Alexander Alexeev, Alireza Sadeghi, and Jonathan
  Howse.
\newblock Synthetic running and tumbling: an autonomous navigation strategy for
  catalytic nanoswimmers.
\newblock {\em Soft Matter}, 8:3077--3082, 2012.

\bibitem{gardiner_b}
Crispin Gardiner.
\newblock {\em Stochastic Methods: A Handbook for the Natural and Social
  Sciences}.
\newblock Springer, 2009.

\bibitem{peruani_prl07}
Fernando Peruani and Luis~G. Morelli.
\newblock Self-propelled particles with fluctuating speed and direction of
  motion in two dimensions.
\newblock {\em Physical Review Letters}, 99:010602, 2007.

\bibitem{romanczuk_prl11}
Pawel Romanczuk and Lutz Schimansky-Geier.
\newblock Brownian motion with active fluctuations.
\newblock {\em Physical Review Letters}, 106:230601, 2011.

\bibitem{pototsky_epl12}
A.~Pototsky and H.~Stark.
\newblock Active brownian particles in two-dimensional traps.
\newblock {\em EPL (Europhysics Letters)}, 98:50004, 2012.

\bibitem{berg_b}
H.C. Berg.
\newblock {\em Random Walks in Biology}.
\newblock Princeton University Press, 1993.

\bibitem{daniels_dev10}
Brian Daniels, Terrence Dobrowsky, Edward Perkins, Sean Sun, and Denis Wirtz.
\newblock Mex-5 enrichment in the c. elegans early embryo mediated by
  differential diffusion.
\newblock {\em Development}, 137:2579--2585, 2010.

\bibitem{griffin_cell11}
Erik Griffin, David Odde, and Geraldine Seydoux.
\newblock Regulation of the mex-5 gradient by a spatially segregated
  kinase/phosphatase cycle.
\newblock {\em Cell}, 146:955--968, 2011.

\bibitem{Zwanzig_b}
Robert Zwanzig.
\newblock {\em Nonequilibrium Statistical Mechanics}.
\newblock Oxford University Press, 2001.

\bibitem{huang_b}
Kerson Huang.
\newblock {\em Statistical Mechanics, 2nd Edition}.
\newblock Wiley, 1987.

\bibitem{lee_pre10}
C.~F. Lee.
\newblock Fluctuation-induced collective motion: A single-particle density
  analysis.
\newblock {\em Physical Review E}, 81:031125, 2010.

\bibitem{BDS}
Brownian dynamics simulation are based on the Langevin equations in Eqs (\ref{eq:rr}-\ref{eq:theta}), and on the corresponding discretised six-direction model (Sect.~5). The dimensionless time increment is set to be 0.01. The dynamics of 1000 particles from random initial conditions are simulated for $5\times 10^6$ time steps. Data presented are collected from the last $2\times 10^6$ time steps in the simulations.

\bibitem{mathematica}
 Wolfram~Research, Inc.
\newblock {\em Mathematica Edition: Version 8.0}.
\newblock Wolfram Research, Inc., Champaign, Illinois, 2010.

\end{thebibliography}

\end{document}